# Dynamics of a single-mode semiconductor laser with incoherent optical feedback

I.V. Koryukin

*Institute of Applied Physics RAS, 46 Ulyanov Street , 603950, Nizhny Novgorod, Russia*

ABSTRACT

A novel model of a semiconductor laser with optical feedback is presented, generalizing Lang-Kobayashi equations to the case of incoherent feedback. The equations are supplemented by a stochastic variable which models random phase difference between the field inside laser cavity and the feedback field. It is shown that for weak-to-moderate feedback the transition from coherent to incoherent feedback leads to replacement of dynamical chaos by almost stationary lasing with slightly fluctuating intensity. Nevertheless, incoherent feedback can lead to chaotic oscillations, but at considerably larger feedback levels.

## I. INTRODUCTION

The recent interest in the dynamics of semiconductor lasers with optical feedback is due to potential applications of such lasers for secure communications by means of chaotic synchronization [1,2]. A solitary semiconductor laser usually displays stable oscillations with steady intensity similarly to other class B lasers, in particular, solid-state lasers. External perturbations, such as pump current modulation, injected signal or feedback are required to achieve a chaotic output. From a practical viewpoint, optical feedback provided by a backreflecting mirror is one of the simplest ways to achieve broadband chaotic oscillations from a semiconductor laser. Even a weak optical feedback leads to complex chaotic regimes [3-5].

The best studied case is that of a coherent optical feedback which corresponds to the situation when the length of the feedback loop is less than the coherence length of laser radiation. In this case, the phase difference between the delayed feedback field and the field inside the laser cavity is constant and the fields interfere. The dynamics of the semiconductor laser with weak-to-moderate coherent optical feedback is adequately described by the well-known Lang-Kobayashi model [6]. The Lang-Kobayashi model is a system of two ordinary differential equations for the slowly varying complex amplitude of the electromagnetic field of laser mode and carrier density (population inversion). Feedback is treated as a single reflection from the external mirror and adds a time delay term in the field equation.

The opposite situation of incoherent feedback is interesting also. It can occur naturally when the length of the feedback loop becomes greater than the coherence length of laser radiation, or can be created specially by rotating the polarization plane of the light in the feedback loop to make it orthogonal to the laser field. In either case there is no interference and the phase difference between the fields has no impact on their interaction which occurs only through joint saturation of the active medium. Of the two cases of incoherent feedback mentioned above, only the case of the polarization plane rotation in the feedback loop was investigated [7-11]. The model adequately describing the experimentally observed dynamics of a semiconductor laser with polarization rotating incoherent feedback is proposed [8]. This model consists of the equation for the field intensity inside laser cavity and the equation for the carrier density. In contrast to the Lang-Kobayashi model, feedback affects only the population inversion, the equation for the field amplitude has no time delay term.

The model of Ref. [8] is not valid in the case of feedback without polarization plane rotation and the feedback loop length exceeding the coherence length of laser radiation. The time-delayed feedback field and the field inside the laser do not interfere, since the phase difference between these fields is not constant but varies randomly. To correctly describe such a situation I consider in this paper a modified Lang-Kobayashi model, in which the phase of the feedback field is a random variable.

## II. MODEL

After appropriate normalization (Ref. [12]) the proposed model can be written in the following dimensionless form:

$$\frac{dE(t)}{dt} = (1+i\alpha)F(t)E(t) + \eta E(t-\tau)\,e^{-i(\Omega\tau+\varphi_r)}$$
$$T\frac{dF(t)}{dt} = P - F(t) - (1+2F(t))|E(t)|^2,$$
(1)

where $E(t)$ is the slowly varying complex amplitude of the electric field of lasing mode, $F(t)$ is the excess free-carrier density, $\alpha$ is the linewidth enhancement factor (Henry's $\alpha$-factor), $\eta$ and $\tau$ are the feedback level and round-trip time of the feedback loop and $\Omega$ is the optical frequency of the solitary laser. The excess pump current $P$ is proportional to $J/J_{th}-1$, where $J$ and $J_{th}$ are the injection current and its value at the solitary laser threshold, respectively. Time $t$ is measured in units of the photon lifetime $\tau_p$, and $T = \tau_s/\tau_p$ where $\tau_s$ is the carrier lifetime.

The stochastic variable $\varphi_r$ is distributed uniformly on the interval $[0...2\pi]$. It introduces into the model random phase difference between the internal and feedback field in the case of incoherent feedback. In the opposite case, when $\varphi_r \equiv 0$, the feedback is fully coherent and model (1) becomes a set of the Lang-Kobayashi equations.

## III. RESULTS OF NUMERICAL SIMULATION

Behavior of the system with coherent and incoherent feedback was compared. Numerical simulation of Eqs.(1) was carried out for two sets of laser parameters, which for coherent feedback lead to different dynamical regimes of laser operation. The first parameter set corresponds to the pump near the solitary laser threshold $P=0.001$. For some feedback range it leads to a complex chaotic regime of low-frequency fluctuations (LFF) with sudden irregular intensity dropouts followed by a gradual intensity recovery [3,4]. In this regime the characteristic frequencies of the intensity modulation are significantly lower than the relaxation oscillation frequency. The second parameter set is chosen well above the threshold $P=0.3$. In this case, typical chaotic behavior is the so-called destabilized relaxation oscillations (DRO) which have irregularities on the time scale of the relaxation oscillations period. Note that both regimes occur when the feedback coefficient exceeds a certain threshold. The fixed parameters for both sets are: τ =3, α=5, T=10³ ($\tau_p$=1 ps, $\tau_s$=1 ns) and Ωτ =0. The feedback coefficient η was varied and used as the control parameter. Calculations of the correlation dimension and spectra were performed with the TISEAN package [13].

For the first parameter set, the difference between the effect of coherent and incoherent feedback is shown in Figure 1. If the feedback is coherent, the laser is in the low-frequency fluctuations regime as depicted in Fig. 1(a),(b). In this regime temporal evolution of the laser intensity has the form of an irregular sequence of packets of picoseconds pulses [Fig. 1(a)]. The typical experimental picture of LFF (intensity dropouts followed by a gradual intensity recovery) is the result of restricted frequency bandwidth of the registration technique. The average laser intensity reproduces well the experimentally observed features of the LFF regime, see Fig.1 (b). Switching on of the phase fluctuations of the feedback field (transition from coherent to incoherent feedback) results in the replacement of chaotic LFF regime by almost stationary lasing with slightly fluctuating (noisy) intensity as shown in Fig. 1(c). Power spectra of the laser intensity for the coherent and incoherent feedback are presented in Figure 2. For coherent feedback the spectrum is typical for the regime of low-frequency fluctuations, it is a set of harmonics of the feedback frequency $f_\tau = 1/\tau \approx 0.33\,\text{GHz}$ [Fig. 2(a)]. In the incoherent feedback case, the power spectrum has the form of low broad peak centered at a relaxation oscillation frequency and does not contain harmonics of the feedback frequency [Fig. 2(b)]. This type of the spectrum confirms a noise nature of the process, the peak is due to excitation of relaxation oscillations by the incoherent feedback signal.

The almost stationary lasing occurs at a relatively small level of incoherent feedback, starting just above the instability threshold of steady-state lasing (η≈0.0002). An increase in η leads to an increase in the intensity fluctuations which, nevertheless, remain random, nondeterministic. A further increase in η leads to 100% modulation of the laser intensity. The calculations of the correlation dimension indicate that the resulting behavior is the dynamical chaos with some additional noise. Typical intensity oscillations at strong incoherent feedback are presented in Fig. 3(b),(d) in comparison with the laser intensity in the case of coherent feedback at the same parameters [Fig. 3(a),(c)]. It is obvious that the chaotic regime in the case of incoherent feedback is not identical to the LFF regime realized at coherent feedback. First, there are no zero dropouts of the intensity envelope, and secondly, filling pulses are longer and rarer.

For the second set of parameters, the effect of the incoherent feedback is similar to the above mentioned LFF case. Chaotic regime of destabilized relaxation oscillations is realized for the coherent feedback as depicted in Fig. 4(a). It has the form of non-harmonic oscillations with frequency close to the relaxation oscillation frequency, with some phase irregularities and chaotic amplitude modulation, see also Fig. 5(c). The power spectrum of this process has the main peak at the relaxation oscillation frequency (3.9 GHz for the chosen parameters) with two asymmetric sets of satellites on its both sides, see Fig. 4(c). As in the previous case, the transition from coherent to incoherent feedback leads to almost stationary lasing with slightly fluctuating intensity as shown in Fig. 4(b). The corresponding power spectrum of the intensity fluctuations has the form of a low broad peak centered at a relaxation oscillation frequency [Fig. 4(d)] as for the first set of parameters, see Fig. 2(b).

For incoherent feedback, an increase in $\eta$ leads to an increase in the intensity fluctuations up to 100% modulation of the intensity, as shown in Fig. 5(b),(d). Calculations of the correlation dimension showed that, when $\eta > 0.1$, the deterministic component dominates in the intensity oscillations, so the system exhibits the regime of dynamical chaos. In contrast to the LFF case, where the chaotic regime at large incoherent feedback is qualitatively different from the chaotic regime at the coherent feedback, in the DRO case these regimes practically coincide (Fig. 5). The only small difference is the presence of a small high-frequency noise in the time series at incoherent feedback [Fig. 5(d)]. The power spectrum of the intensity at strong incoherent feedback differs fundamentally from that of the case of weak incoherent feedback (Fig. 6), it is qualitatively the same as the spectrum of destabilized relaxation oscillations [Fig. 4(c)].

## IV. CONCLUSION

For both LFF and DRO regimes, switching on of the phase fluctuations of the feedback field (transition from coherent to incoherent feedback) leads to the replacement of dynamical chaos by almost stationary lasing with noisy intensity. This behavior occurs at a relatively small feedback level, just above the threshold of the corresponding chaotic regime. At incoherent feedback, an increase in $\eta$ leads to an increase in the intensity fluctuations which, nevertheless, remain essentially random, nondeterministic. A further increase in $\eta$ leads to 100% modulation of the laser intensity. At strong incoherent feedback, the system exhibits the regime of dynamical chaos with some additional noise.

Thus, even incoherent optical feedback can lead to chaotic oscillations of the laser intensity. However, unlike the case of coherent feedback, chaotic oscillations are achieved at considerably larger feedback levels. In addition, the instability of the stationary lasing has no threshold behavior, or its threshold is very blurry. Moreover, in the case of incoherent feedback, the proposed model can have both noise and deterministic chaotic solutions. This means that it is impossible to build a purely dynamic model of the semiconductor laser with incoherent feedback applicable in the whole range of parameters.

ACKNOWLEDGMENTS

This research was supported by the Russian President Program for Support of Leading Scientific Schools (Grant No. 5430.2012.2).

FIGURE CAPTIONS

Fig. 1. Time-dependent solution of model (1) for $P = 10^{-3}$, $\eta=0.01$.
(a) Laser intensity and (b) averaged laser intensity for coherent feedback (LFF), (c) laser intensity for incoherent feedback. Averaging time for plot (b) is 2 ns.

Fig. 2. Power spectral density of the laser intensity for the parameters of Fig.1. (a) Coherent feedback (corresponding to Fig.1a), (b) incoherent feedback (Fig.1c).

Fig. 3. Time-dependent solution of model (1) for $P = 10^{-3}$, $\eta=0.25$. Laser intensity (a,c) for coherent feedback and (b,d) for incoherent feedback.

Fig. 4. Time-dependent solution of model (1) for $P = 0.3$, $\eta=0.005$.
Laser intensity and corresponding power spectral density (a,c) for coherent feedback (DRO) and (b,d) for incoherent feedback.

Fig. 5. Laser intensity (a,c) for weak incoherent feedback $\eta=0.01$, (b,d) for strong incoherent feedback $\eta=0.15$. $P = 0.3$.

Fig. 6. Power spectral density of the laser intensity (a) for weak and (b) for strong incoherent feedback. $P = 0.3$, (a) $\eta = 0.01$, (b) $\eta = 0.15$.

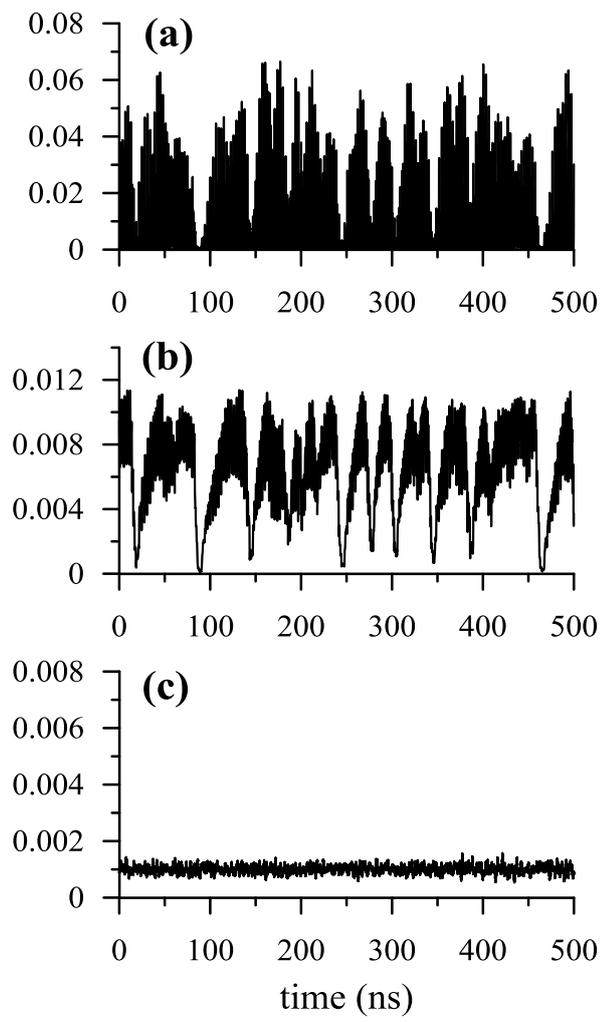

Fig.1

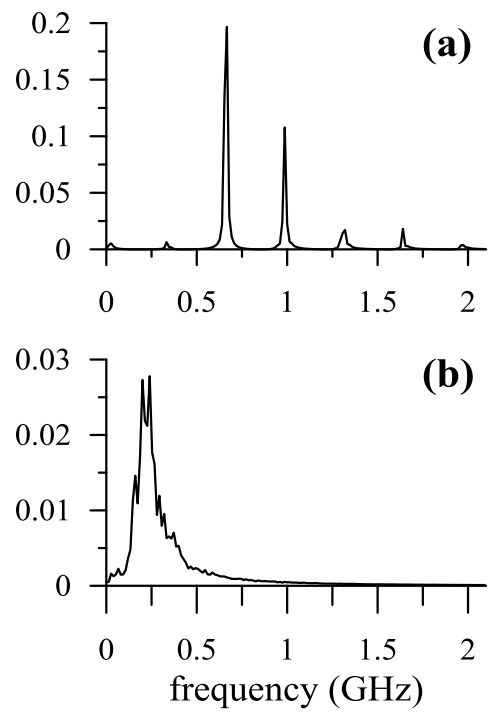

Fig.2

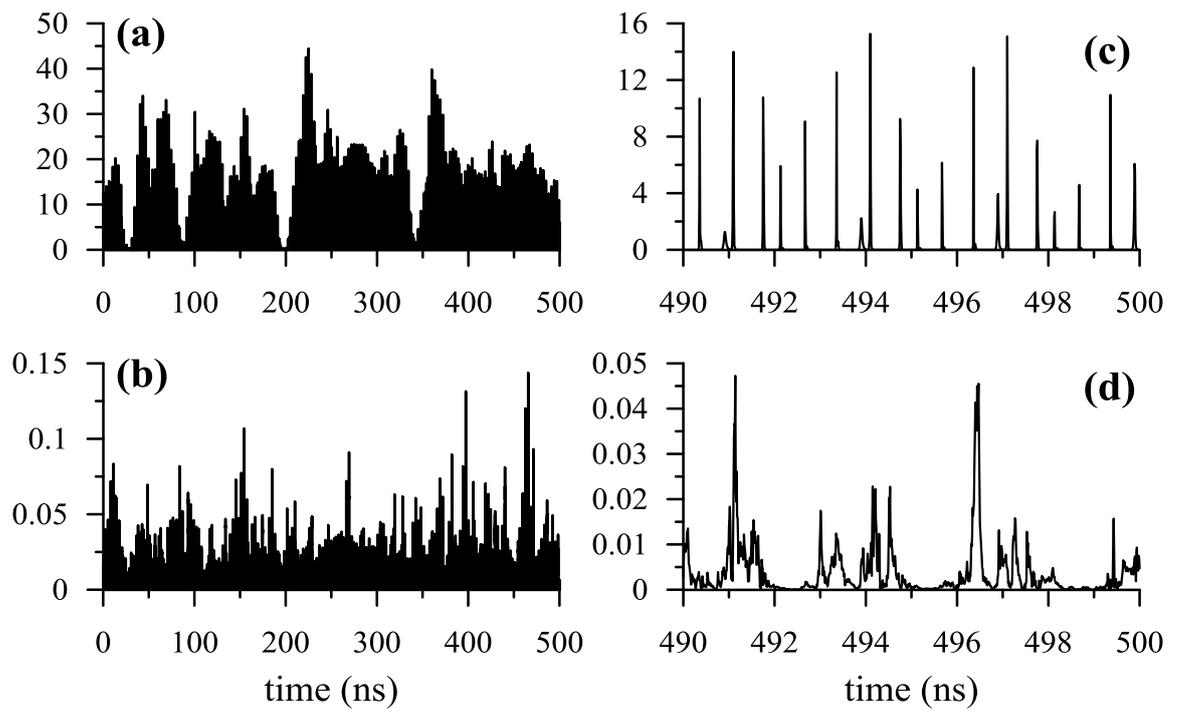

Fig.3

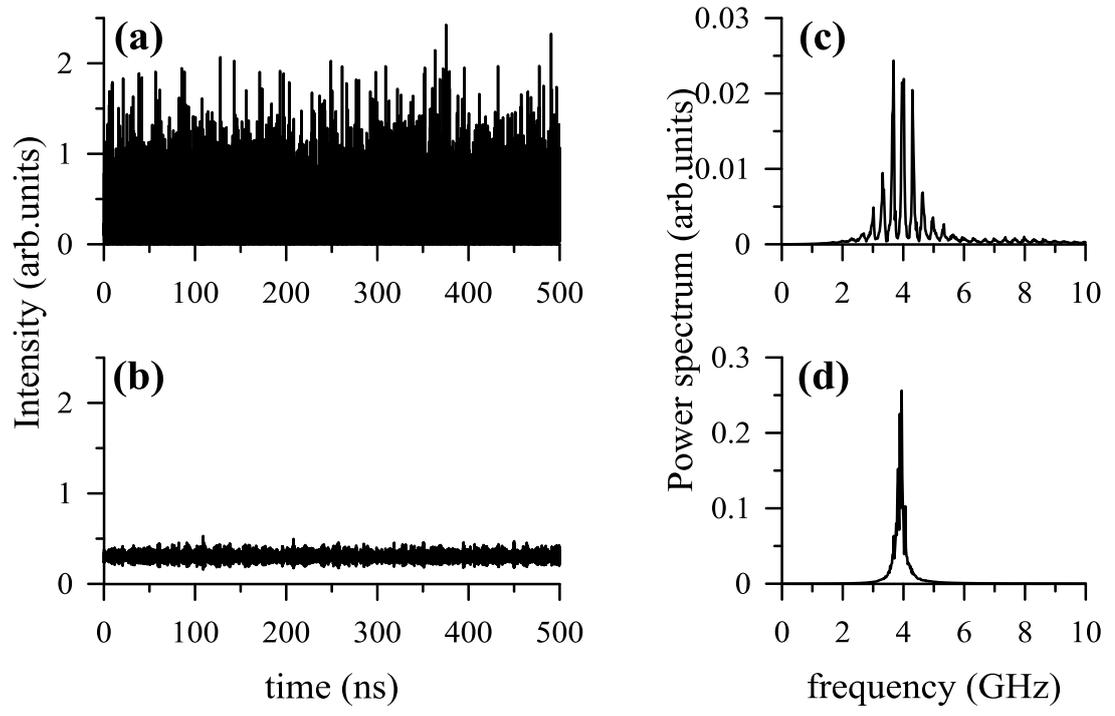

Fig.4

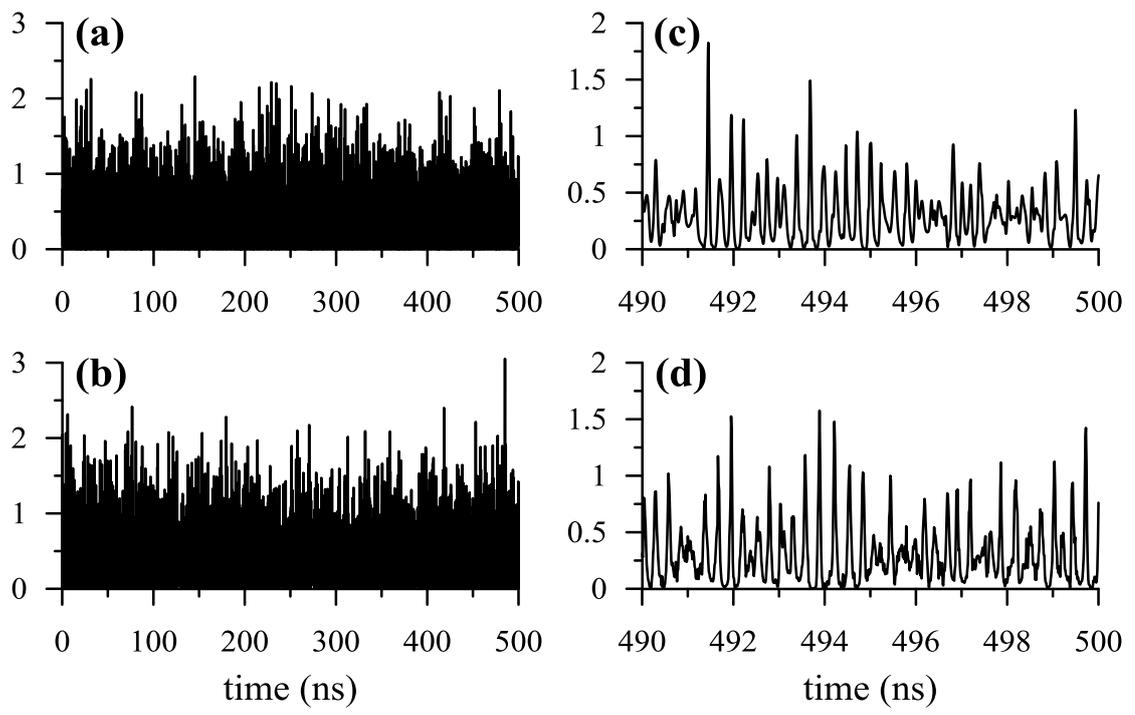

Fig.5

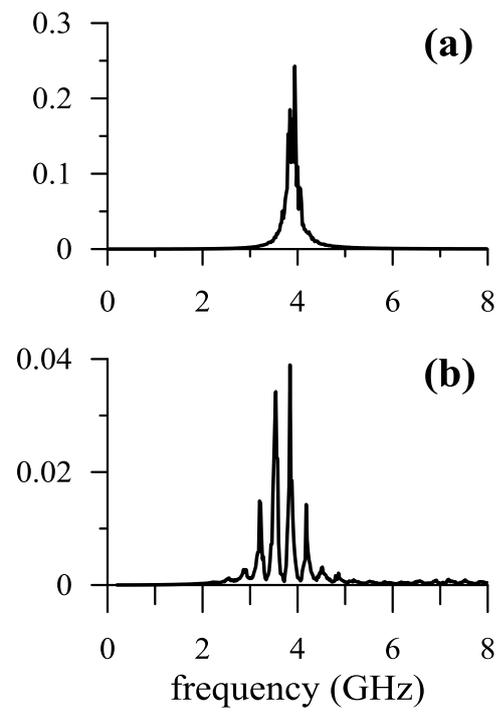

Fig.6